\documentclass[11pt,aps,twocolumn,superscriptaddress,nofootinbib,tightenlines,floatfix,longbibliography]{revtex4-1}
\usepackage{geometry} % to change the page dimensions
\usepackage{amsmath,amssymb}
\usepackage{graphicx}
\usepackage{color}
\usepackage{comment}
\usepackage{bbold}
\usepackage{soul,color}
\sethlcolor{white}
\usepackage{booktabs}
\usepackage{array}
\usepackage{hyperref}
\hypersetup{
    colorlinks=true,       % false: boxed links; true: colored links
    linkcolor=blue,          % color of internal links
    citecolor=blue,        % color of links to bibliography
    filecolor=blue,      % color of file links
    urlcolor=blue           % color of external links
}

\begin{document}

\title{How about that Bayes: Bayesian techniques and the simple pendulum}

\author{Matthew Heffernan}
\email[]{heffernan@physics.mcgill.ca}

\affiliation{Department of Physics, 
	McGill University, 3600 University Street, Montreal, QC, H3A 2T8, Canada}

\begin{abstract}
    Physics increasingly uses Bayesian techniques for systematic data analysis and model-to-data comparison. This paper describes how these methods can be implemented to answer questions of relevance to teaching laboratories. It demonstrates the Bayesian approach to statistical modeling and model selection in a step-by-step workflow. The simple pendulum provides a demonstration with the precision commonly seen in the introductory laboratory. This is used to provide realistic, quantitative guidance for model preference between the small angle approximation and more complicated formula. This extends the simple pendulum literature's focus beyond comparing individual idealized assessments of different approximations and provides actionable, data-driven guidance for teaching laboratory design.
\end{abstract}

\maketitle

\section{Introduction}

    Introductory physics courses commonly teach that the period of a simple pendulum displaced by a small angle $\theta$ may be found by applying the approximation $\sin(\theta) \approx \theta$. The period of a pendulum also depends on the initial angular displacement. Attempts to increase accuracy in approximations of the pendulum period have focused on more closely approximating the full non-analytic integral expression for the period so students in undergraduate laboratories can investigate the dependence of the period on initial angular displacement \cite{brazilapprox}. 
    Recently, model comparison and complexity has been applied to the pendulum problem in the teaching laboratory \cite{Modelcomparison}.
    
    Guidance for the use of the small angle approximation in simple pendula has not been rigorously quantified using realistic uncertainties. Often, the only suggestion is that after $\sim 15^\circ$ the difference between $\sin(\theta)$ and $\theta$ in radians exceeds 1\% and the small angle approximation should no longer be used \cite{giancoli2016physics}. This does not take into account measurement uncertainties found in a laboratory setting or if the data itself requires additional complexity in the form of more elaborate formulae; c.f. \cite{Hinrichsen_2020} for a recent review. This study demonstrates that students with both an exact and analytical formula for the period of a pendulum can successfully establish a quantitative preference between models at moderate displacements using Bayesian tools. 
    This study remedies a gap in the literature by quantitatively motivating more complex formula in teaching laboratories via realistic measurements. 
    
    This paper demonstrates how to establish quantitative criteria for the breakdown of the small angle approximation by introducing and demonstrating the techniques of Bayesian data analysis. Using Bayesian model selection, a quantitative preference between the small angle approximation and a numerical calculation of the exact integral equation is determined. First, an introduction to Bayesian statistics is provided. The exact expression for the period of a simple pendulum -- without nonconservative forces, such as drag or friction -- is derived as well as the small angle  approximation to establish the Physics background. It is demonstrated that it is possible to differentiate the two formula with realistic measurement uncertainties. This is followed by using Bayesian inference to compare the models to data; first to pseudodata generated by each individual model to determine self-consistency and then to measurements of the period of a simple pendulum. 
    
    The posterior distributions are then analyzed and posterior predictive distributions are consistent with expectations. Finally, it is determined at what initial angular displacement weak, moderate, and strong preference for the exact calculation are found given measurements with a variety of timing uncertainties. This preference should be used to inform recommendations of maximum initial angular displacement in laboratory materials and textbooks. The overview and workflow presented are intended as a general introduction to Bayesian techniques for implementation in the undergraduate physics curriculum.

\section{Bayesian Statistics}

    Bayesian data analysis is an increasingly-commonly used tool in physics research, \textit{e.g.} gravitational-wave astronomy \cite{10.1093/mnras/staa2483} and recently in ultra-relativistic heavy ion collisions \cite{SIMS}, among many others \cite{bayesinphysics}. The techniques and process are ripe for application to pedagogical questions. 
    
    In Bayesian statistics, probability is interpreted as the degree of belief in a proposition given the available data and prior knowledge. Put another way, Bayesian statistics quantifies the plausibility or betting odds of a proposition being true. The Bayesian approach is well-suited to applications with limited data in which other knowledge may exist -- a description that fits the introductory laboratory well. 
    
    The central theorem of Bayesian statistics is Bayes' theorem,
    \begin{equation}
	    p(H|d,I) = \frac{p(d|H,I)\hspace{0.1in}p(H|I)}{p(d|I)}. \label{eq:bayes}
	\end{equation}
	$p(\cdot)$ denotes probability density and the vertical bar in $p(\cdot|\cdot)$ denotes conditionality, \emph{i.e.} $p(A|B, C)$ is the probability density of $A$ given $B$ and $C$. A colloquial description of Bayes' theorem is that it formalizes the process of learning: a prior belief is compared with data. Based on how well it matches the observation, the prior belief is determined to be relatively more or less likely. This updated understanding forms the posterior belief. $H$ denotes a hypothesis such as a particular choice of parameter value and $I$ denotes any other information, such as a particular model. In practice, the hypothesis $H$ often represents a particular choice or distribution of parameter(s). $I$ may also denote any other additional information such as reasonable expectations that are not directly informed by $d$, such as expectations from first-principles theory. $p(H|I)$ is called the ``prior" and represents the belief in the hypothesis $H$ \emph{prior} to comparison with measurements. $p(d|H,I)$ is the ``likelihood" and is the likelihood for the data $d$ to be true given the hypothesis $H$. $p(d|I)$ is the ``Bayes evidence" and is typically treated as a normalization constant such that $\int p(H|d,I) dH =1$. The Bayes evidence quantifies how likely one believes the data to be given other information $I$. It can therefore be used in model selection: as the data must be true, the model that finds the data most likely is the model preferred by the data.
	$p(H|d,I)$ is called the ``posterior" and quantifies the belief in any hypothesis $H$ \emph{posterior} to comparison with measured data $d$.
	
	Bayes' theorem is used to solve inverse problems, calculating ``how likely is the data given a set of model parameters" (the likelihood) in order to answer a primary question of interest, ``how likely is the set of model parameters given the data" (the posterior). Consequently, a hypothesis for which the observed data is unlikely is itself an unlikely hypothesis. These concepts will be demonstrated using the familiar case of the simple pendulum. 
	
	Bayes' Theorem is typically evaluated using computational tools such as Markov Chain Monte Carlo (MCMC). This yields an estimate of the parameters of the model as well as any covariance structure. An introduction to Bayesian methods, including contrasts with Frequentist statistics, can be found in \cite{BayesintheSky,sivia2006data,gelman2013bayesian}. A thorough overview of Bayesian modeling may be found in \cite{gelman2013bayesian,gelman2020bayesian, betancourtfalling}.
	
	In this study, Bayesian techniques are used to determine when the data exhibits weak, moderate, and strong preference for the exact expression for the period of a pendulum relative to the expression for the period of a pendulum in the small angle approximation. To do so, realistically-generated pseudodata and motivated prior expectations will be used in order to evaluate Bayes' Theorem (Eq.~\ref{eq:bayes}) in a modeling workflow. This study employs the use of the Gaussian likelihood function, usually represented in its logarithmic form, which assumes that uncertainties are normally distributed around a central value.
	\begin{equation}
	    \ln p(d|H,I) =  -\frac{1}{2} \sum_{i=1}^N \left[ \ln(2\pi  \sigma^2) + \frac{(d_i - Y_{\rm Model, i})^2}{\sigma^2_i} \right],\label{eq:likelihood}
	\end{equation}
	where $Y_{\rm Model, i}$ is the model calculation of the period at the $i^{th}$ of $N$ displacements. The variance $\sigma^2_i = \sigma_{\rm data_i}^2 + \sigma_{\rm model_i}^2$ is the data uncertainty and the model uncertainty added in quadrature. Model uncertainty arises from uncertain inputs; in this study the length of a pendulum as well as the angular displacement are both measured inputs and therefore have associated uncertainty. Only one parameter -- the gravitational acceleration $g$ -- is varied in the inference and a particular value of $g$ constitutes a hypothesis $H$ given the choice of model or other assumptions $I$. Uncertainty for each model is propagated from uncertain inputs assuming that the variables are independent (or uncorrelated),  where for a function $f$ of variables $a$, $b$, etc., the variance of the model prediction is given by
	\begin{equation} 
	\sigma_f^2 = \left(\frac{\partial f}{\partial a}\right)^2\sigma_a^2 + \left(\frac{\partial f}{\partial b}\right)^2\sigma_b^2 + \dots,\label{eq:variance}
	\end{equation}
	and is known as the variance formula and discussed in classic works, \emph{c.f.}  \cite{ku1966notes} and references therein. Generalizing to higher dimensions is beyond the scope of this work, but is well documented in standard textbooks \cite{sivia2006data,gelman2013bayesian}.
	
	\section{The period of a simple pendulum}
	
    An ideal simple pendulum has a bob of mass $m$ suspended from a frictionless attachment point by a massless rod of length $L$ and moves in a single angular dimension $\theta$. In the absence of non-conservative forces such as friction and drag, the equation of motion is
    \begin{equation}
        \ddot{\theta} = - \frac{g\sin (\theta)}{L} \label{eq:ddottheta}
    \end{equation}
    where $g$ is the gravitational acceleration. 
    For small displacements, $\sin{\theta}\approx \theta$ and Eq.~\ref{eq:ddottheta} can be identified as a simple harmonic oscillator. The period of a simple harmonic oscillator is $T = 2\pi \sqrt{\frac{m}{k}}$ and it is trivial to identify $k=\frac{mg}{L}$. Consequently, 
    \begin{equation}
        T_0 = 2\pi \sqrt{\frac{L}{g}} \label{eq:sa}
    \end{equation}
    and the uncertainty on the angular measurement does not enter into the model calculation while the uncertainty on the length does. 
    
    An exact expression for the period of a simple pendulum is readily derived using the conservation of energy. The expression is simplified by choosing the zero of potential energy when $\theta(t)=0$ and assuming the pendulum to be initially stationary and at its initial angular displacement $\theta_0$. The energy conservation equation is
    \[
        m g L\left(1-\cos \theta_{0}\right)=\frac{1}{2} m L^{2}\dot{\theta}^{2}+m g L(1-\cos \theta),
    \]
    again neglecting non-conservative forces. 
    It is then straightforward to solve for $\dot{\theta}$ and
    integrate $\theta$ from 0 to $\theta_0$, corresponding to one-quarter of the period. This yields
        $T=2 \sqrt{2} \sqrt{\frac{L}{g}} \int_{0}^{\theta_{0}} \frac{1}{\sqrt{\cos \theta-\cos \theta_{0}}} d \theta.$
    This integral is improper when $\theta=\theta_0$, but a simple substitution $\cos \theta = 1-2\sin^2(\theta/2)$ may be combined with a change of variables $\sin \phi = \frac{\sin(\theta/2)}{\sin(\theta_0/2)}$ to yield
    \begin{equation}
        T=4\sqrt{\frac{L}{g}} \int_{0}^{\pi/2} \frac{1}{\sqrt{1-\sin^2(\theta_0/2) \sin^2(\phi)}} d \phi. \label{eq:exact}
    \end{equation}
    As this expression incorporates the initial angular displacement $\theta_0$, the uncertainty on the measured initial angular displacement must be incorporated. Uncertainty on $\theta_0$ is propagated through the integral expression in a technical calculation with the assumption that uncertainties are normally distributed and uncorrelated, using Eq.~\ref{eq:variance}. The exact expression also must account for the measured uncertainty on the length of the pendulum $L$.
    The small angle approximation (Eq.~\ref{eq:sa}) and the exact expression (Eq.~\ref{eq:exact}) are the models considered in this work.
    
	Using real measurements in order to determine model preference thresholds demands a sufficiently large number of measurements as to be impractical\footnote{For 5 measurements of 10 periods at 20 displacements each with 3 timers, a single maximum angular displacement can require 300 measurements.}, so the data used in this study for model selection is generated using Eq.~\ref{eq:exact}. 
	
	Generated data used in model validation are termed ``pseudodata", real measurements are performed for the inference of $g$, and generated data are used to calculate model preference thresholds.
	Pseudodata are generated for model validation by making a set of simple choices informed by experience in undergraduate teaching laboratories. In these laboratories, standard advice about the uncertainty is that the uncertainty of a measurement is equal to half of the smallest increment of the measurement device. A common measuring device in teaching laboratories is a meter stick with millimeter gradations, corresponding to an uncertainty of $\pm 0.0005$ m. 
	
	In this study, a fixed value $L = 0.807 \pm 0.0005 \text{m}$ was measured. The uncertainty on initial angular displacement corresponds to the use of digital protractor with 0.01$^\circ$ gradations; all angular model inputs thus have uncertainty $\pm$0.005$^\circ$. 
	The pendulum length and initial angular displacement are measured and have associated measurement uncertainties that are consistently accounted for in the likelihood function via uncertainty propagation (Eq.~\ref{eq:variance}) in a way that is standard in both Bayesian and Frequentist inference.
	
	In teaching laboratories, more precise timing mechanisms may also be available and the use of smartphone accelerometers in teaching laboratories has become more common \cite{McGill101}. These measurements are significantly more precise than the precision of a stopwatch and are not impacted by human reaction times. Common measurements with smartphone accelerometers in recent courses at McGill University had uncertainty of approximately $\pm 0.02$ s \cite{McGill101}, while timing uncertainties with a photodiode timer used in the measurements performed for this work are smaller still -- $\pm 0.005$ s. Finally, timing uncertainty corresponding to the use of a stopwatch in model selection is taken to be $0.250\sqrt{2}$ s, corresponding to a reaction time of $0.250$ s \cite{10.1093/geronj/44.2.P29} to start and stop the timer added in quadrature. Advice given to students is to time up to 10 periods of a pendulum in order to reduce the relative error on the measurement, a process replicated in this work. This study assumes that measurements are normally distributed with standard deviation corresponding to the uncertainty. 
	
	Different values of the timing precision may be specified when generating pseudodata; it is then possible to specify the initial angular displacement and produce pseudodata that reflects the precision of data in introductory teaching laboratories. 

\section{A workflow for reproducible inference}

    Reliable, reproducible, and robust data analysis is supported by a well-designed workflow. Following a step-by-step process acts as a safeguard that supports a rich understanding or straightforward troubleshooting of the models. 
    In both research and pedagogical settings, clearly-defined steps are able to guide analysis and allow for careful implementation and refinement. An additional benefit of a modeling workflow for teaching applications is that it allows for the analysis to be broken into digestible, concrete steps in course materials.
    
    The workflow presented here progresses from explicitly defining prior knowledge all the way to verifying that the final inference reproduces expectations. Each step is described, motivated, and implemented in the models before progressing to the next step. At the end of the workflow, Bayesian model selection is introduced to establish rigorous criteria for when the data itself demands one model over the other.

	\subsection{Step 1: Defining the prior state of knowledge}
	\label{sec:prior}
	
    Once the models have been chosen, the first step in Bayesian data analysis is to clearly motivate the choice of prior. The prior should reflect concrete understanding of the underlying quantity and be sufficiently general as to constrain the problem without pre-determining or biasing the result \cite{stanpriors}. 
    For example, a prior that precludes the true result can never recover the true result as is clear from Bayes' Theorem. 
    The goal of the prior is to be constraining, but allow for the bulk of the posterior constraint to be determined by systematic comparison of the model to data via the likelihood.
    
    The initial angular displacement and period of a simple pendulum are easily measured and the length of the pendulum has been fixed. As a result, the only free model parameter in both the equations using the small angle approximation and the exact expression for the period of a pendulum is the gravitational acceleration $g$. 
    $g$ must be positive definite and nonzero, otherwise no oscillation will occur.
    It is also possible to ascribe physical intuition to the problem, which will be done with an exaggerated example. 
    Gravitational acceleration on the surface of a body scales with the mass of the body and gravitational acceleration on the Moon is approximately 1.625 $m/s^2$ \cite{moongravity} and the surface gravity of Jupiter is 24.79 $m/s^2$ \cite{jupitergravity}. Proceeding with confidence that the mass of the Earth is likely between that of the Moon and Jupiter, a weakly-informative prior can be constructed that allows for some probability of surprise. 
    
\begin{figure}[!htbp]
        \centering
        \includegraphics[width=\columnwidth]{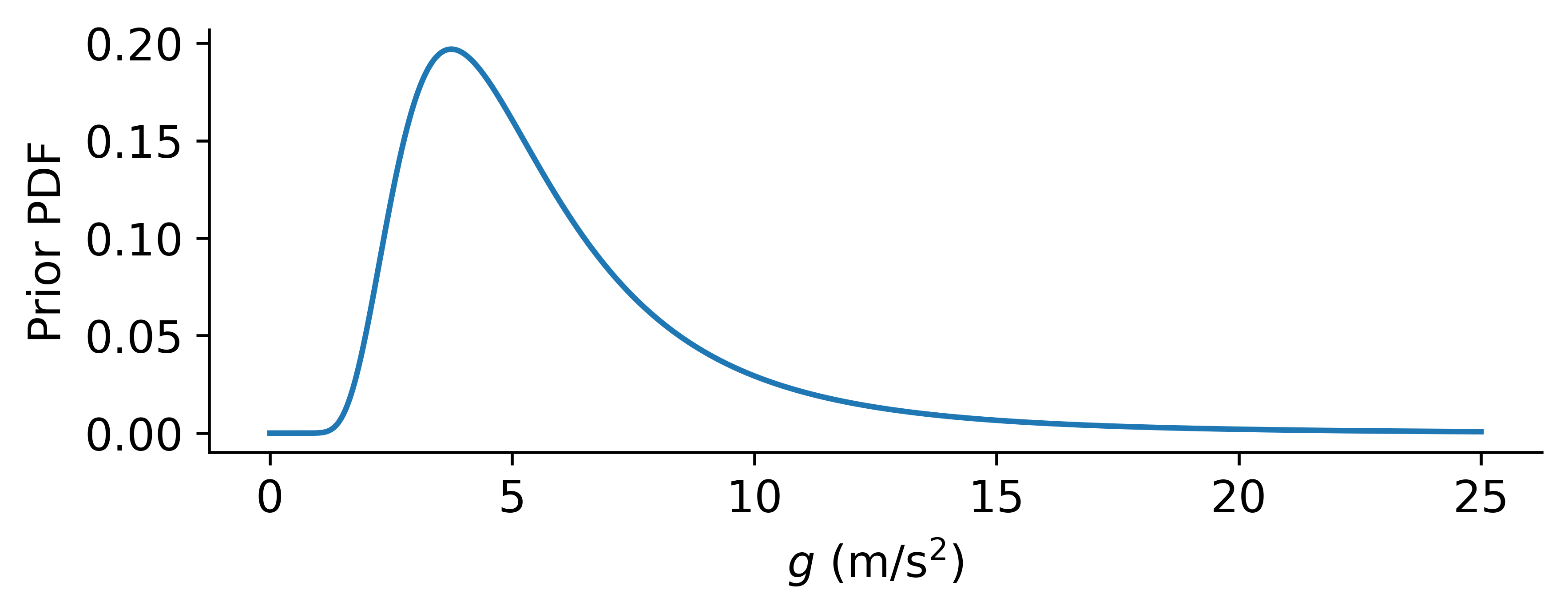}
        \caption{(Color online) The prior probability density function (PDF) of the gravitational acceleration $g$. This PDF is $p(H|I)$ in Eq.~\ref{eq:bayes} where each value of $g$ is a hypothesis $H$.}
        \label{fig:g-prior}
    \end{figure}

    A common choice for a weakly-informative prior on positive definite quantities is the inverse gamma distribution. The inverse gamma distribution has support on the interval $(0,\infty)$, matching the positive definite constraint and is easily tuned so that only 1\% of the probability falls below 1.625 $m/s^2$ and above 24.79 $m/s^2$. 
    This prior is shown in Fig.~\ref{fig:g-prior}. The location of the mode of the distribution is not particularly relevant -- by constructing a weakly-informative prior, the information contained in the data will far outweigh the information in the prior. Other choices of prior that have similar support are possible, but the inverse gamma distribution smoothly approaches 0 on the bounds of its support. This matches physical intuition. For example, if the prior was the uniform distribution $U(1.625,24.79)$, this would not change the results of the inference, but would require justifying that 24.8 $m/s^2$ was completely forbidden while 24.79 $m/s^2$ was allowed and just as likely as any other value in the prior. 
    
	\subsection{Step 2: Prior predictive checks}
	
\begin{figure}[!htbp]
        \centering
        \includegraphics[width=\columnwidth]{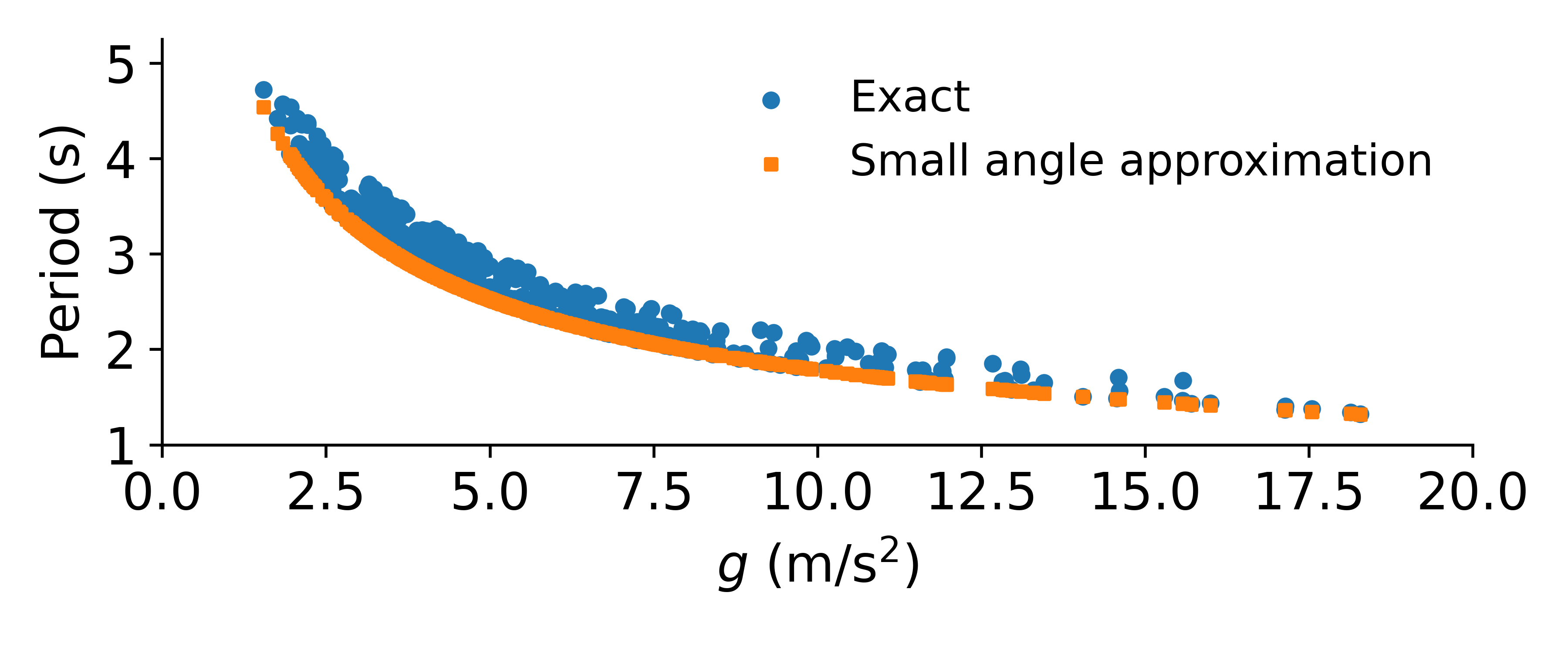}
        \caption{(Color online) 100 predictions for the period with $g$ sampled from the prior and initial angular displacement $\theta_0$ sampled from a uniform distribution $U(0,\pi/4)$.}
        \label{fig:g-prior-pred}
    \end{figure}

	In Physics, limiting-case arguments are a classic example of how to use exploration to set constraints. 
	Prior predictive checking is another example of exploring the parameter space to gain an understanding of model behavior \cite{Box1980SamplingAB}. 
	The computational model is evaluated with draws from the prior distribution to form the \emph{prior predictive distribution}. 
	This is used to assess the underlying model and determine the presence of non-trivial features or predictions that are at odds with expectations. In this simple case, a prior defined with support $[0,\infty)$ would suggest that a finite value for the period should be expected when the gravitational acceleration is 0. However, the period is ill-defined at 0 m/s$^2$ and the result is at odds with expectations. 
	The models in this study are simple and so few unexpected features exist, but the step is included to demonstrate modeling procedure. Predictions for the period with $g$ sampled from the prior distribution are shown in Fig.~\ref{fig:g-prior-pred}. 
	It is expected that measured periods of the pendulum in this work are between approximately 1 and 5 seconds, while longer and shorter periods are possible, but implausible. 
	
	\subsection{Step 3: Model validation}
	
	Model validation, also known as closure testing or empirical coverage, is a means of assessing self-consistency. This is established by solving the statistical inverse problem -- given some output, what is the corresponding input -- with Bayes' Theorem. Pseudodata are generated with each model and the model that generated the data is used in inference to see if the most likely parameter values match the known inputs. This is an important step as it represents the best-case scenario: the model is known to contain all the features of the data. %

\begin{figure}[!htbp]
        \centering
        \includegraphics[width=\columnwidth]{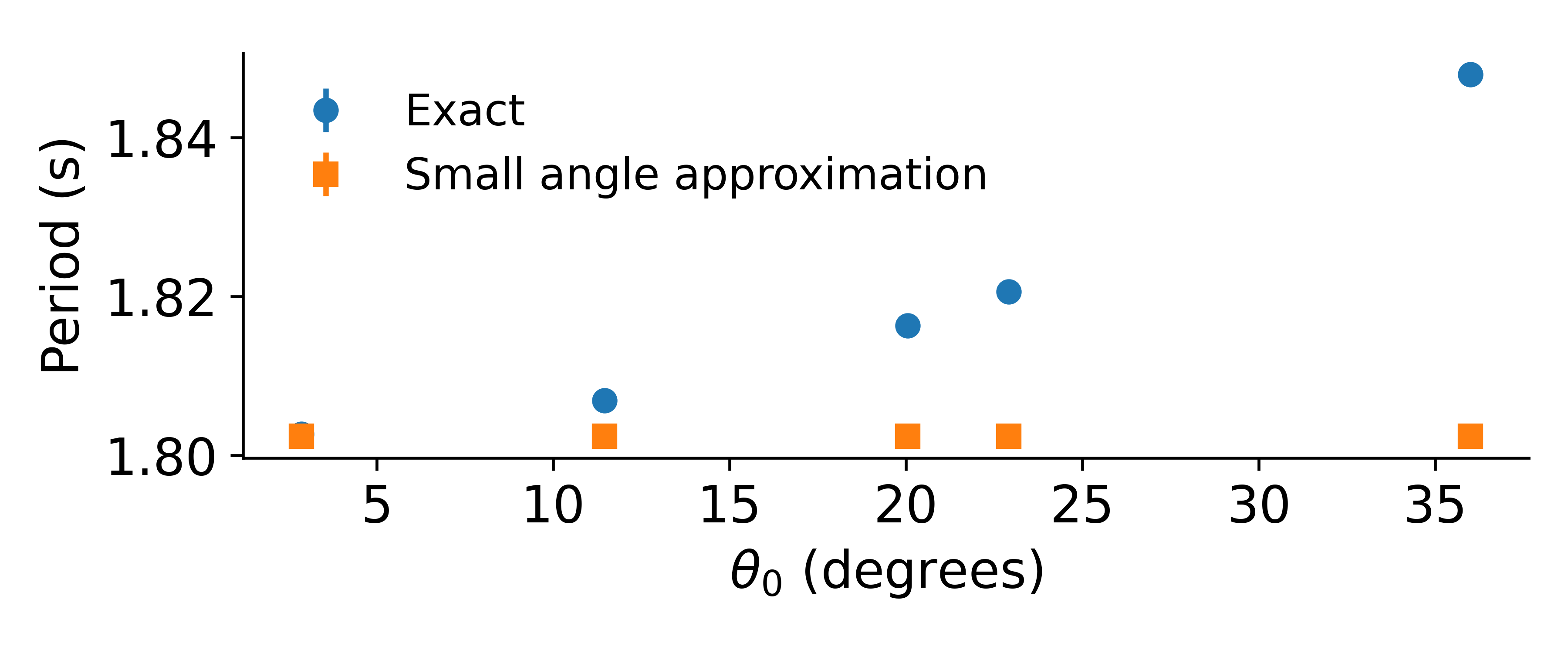}
        \caption{(Color online) Pseudodata used for model validation. In model validation, models are compared to data generated by the model under consideration.}
        \label{fig:closure-data}
    \end{figure}

	Pseudodata are generated with the NIST reference $g = 9.80665$ $ m/s^2$ \cite{NIST_g} and timing uncertainty corresponding to the use of a photodiode timer over 10 periods, comparable to the measurements in the inference with data. Initial displacements are chosen with arbitrary spacing to exaggerate features of real measurements rather than provide an unrealistically-even test case, 
	$\theta_0 = \left[2.86, 11.46, 20.05, 22.92, 35.98 \right]$ degrees, 
	and calculations of the corresponding periods are shown in Fig.~\ref{fig:closure-data}. 
    
\begin{figure}[!htbp]
        \centering
        \includegraphics[width=\columnwidth]{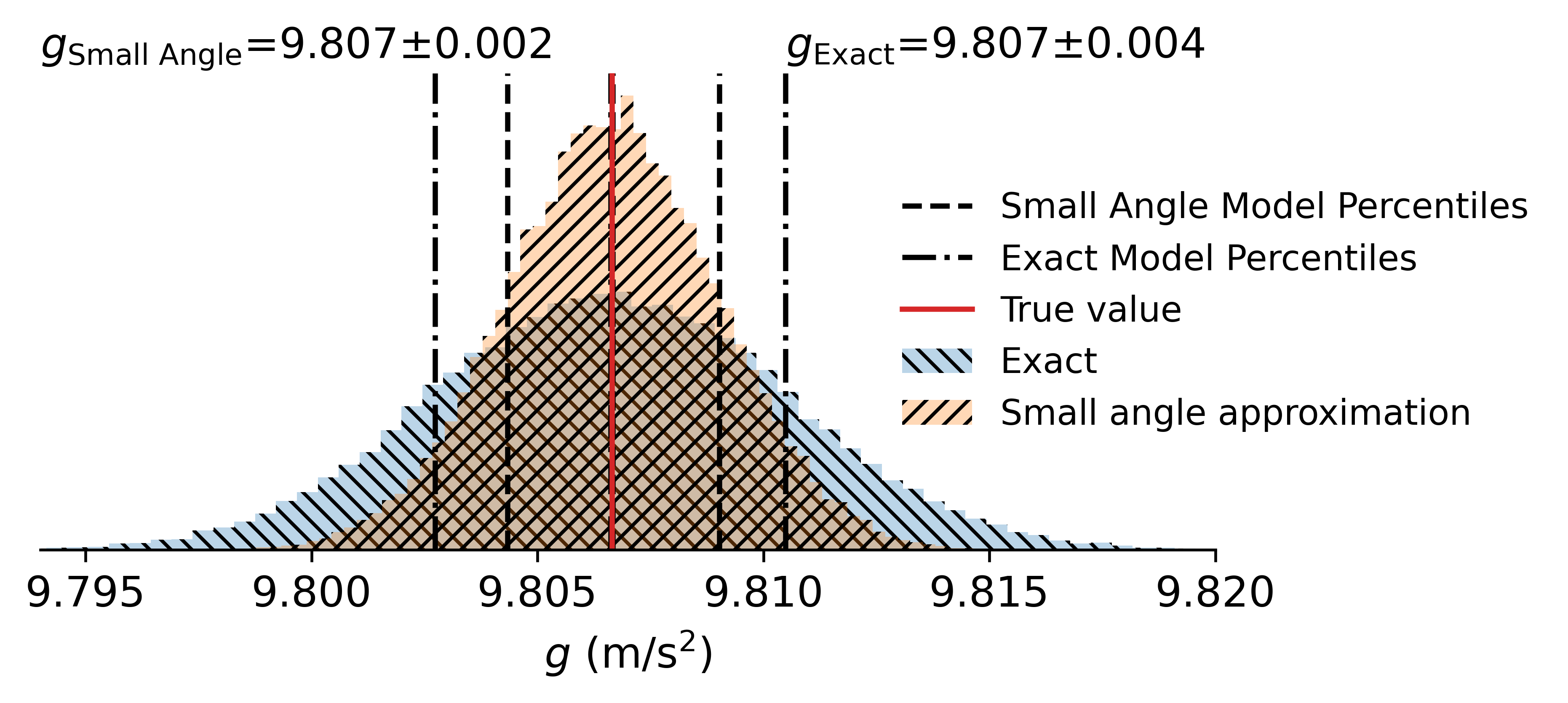}
        \caption{(Color online) Posterior distributions for $g$ in model validation using the exact formula for the period and the small angle approximation. The true value is shown in red and vertical dashed lines denote the 16th, 50th, and 84th percentile of the samples in increasing $g$ with exact values provided above the figure. When the models contain the full information present in the data to which they are compared, both are able to recover the true value. }
        \label{fig:closureposteriors}
    \end{figure}

    The models are both calibrated to their respective validation pseudodata and the posterior distributions are shown in Fig.~\ref{fig:closureposteriors}. It can be clearly seen that both models are able to achieve closure when the data is perfectly consistent with the model. 
    The small angle approximation model's posterior is slightly more peaked than that of the exact model; this is because the pseudodata is more precise. This precision arises because the small angle approximation does not account for uncertainty on $\theta_0$ and thus does not propagate this uncertainty to the calculated period, while the exact expression must carefully account for this additional uncertainty. 
    For each parameter value considered, the likelihood is calculated and the value of the posterior is computed. While this is simple in a small number of dimensions, this is impractical for high-dimensional applications and computing the posterior directly becomes computationally expensive. 
    
    To efficiently compute the posterior, Markov Chain Monte Carlo (MCMC) is used. MCMC algorithms are efficient sampling techniques in which ``walkers" walk through the parameter space according to an algorithm. At each MCMC step, the likelihood is calculated and multiplied by the prior and used to determine if a step is accepted (added to the ensemble of samples) or not. Therefore, each step taken by a walker is an accepted sample from the posterior and the ensemble of samples can be used to estimate properties of the true posterior. The longer the walkers walk, the more samples are drawn and the distribution of those samples more closely resembles the target distribution. These samples can then be analyzed as draws from the posterior. This study uses a numerical implementation of MCMC, \texttt{ptemcee}, whose features allow for straightforward calculation of the Bayes evidence \cite{ptemcee, ptemcee-code, emcee}.  
    \iffalse
    The MCMC behavior must be checked to ensure that the chains converged to their target distributions, \textit{i.e.} that samples are indeed draws from the distribution of interest. One way to check this is by considering autocorrelation, which quantifies how correlated samples are within a certain ``lag".\footnote{Autocorrelation is calculated by taking the inner product of the unshifted chain of samples with a chain shifted by the lag, normalized by the product of the magnitudes of the shifted and unshifted chains. As a result, the 0 lag autocorrelation is 1 by construction.} Independent samples of the posterior should be uncorrelated, while correlated samples suggest that the walkers are walking in a particular direction and do not represent draws from the distribution of interest. The autocorrelation for the exact and small angle approximation MCMC chains are shown in Fig.~\ref{fig:closureautocors}, which show that after even a single step, the chain rapidly decorrelates and is sampling the target distribution successfully. 
    \begin{figure}[!htbp]
        \centering
        \includegraphics[width=\columnwidth]{./figs/autocors_preciseclosure}
        \caption{(Color online) Autocorrelation of MCMC chains for model validation. The rapid drop and subsequent small-scale fluctuations around 0 are indicative of a decorrelated chain, meaning that the chain is sampling from the target distribution (the posterior).}
        \label{fig:closureautocors}
    \end{figure}
    \fi 
    
\begin{figure}[!htbp]
        \centering
        \includegraphics[width=\columnwidth]{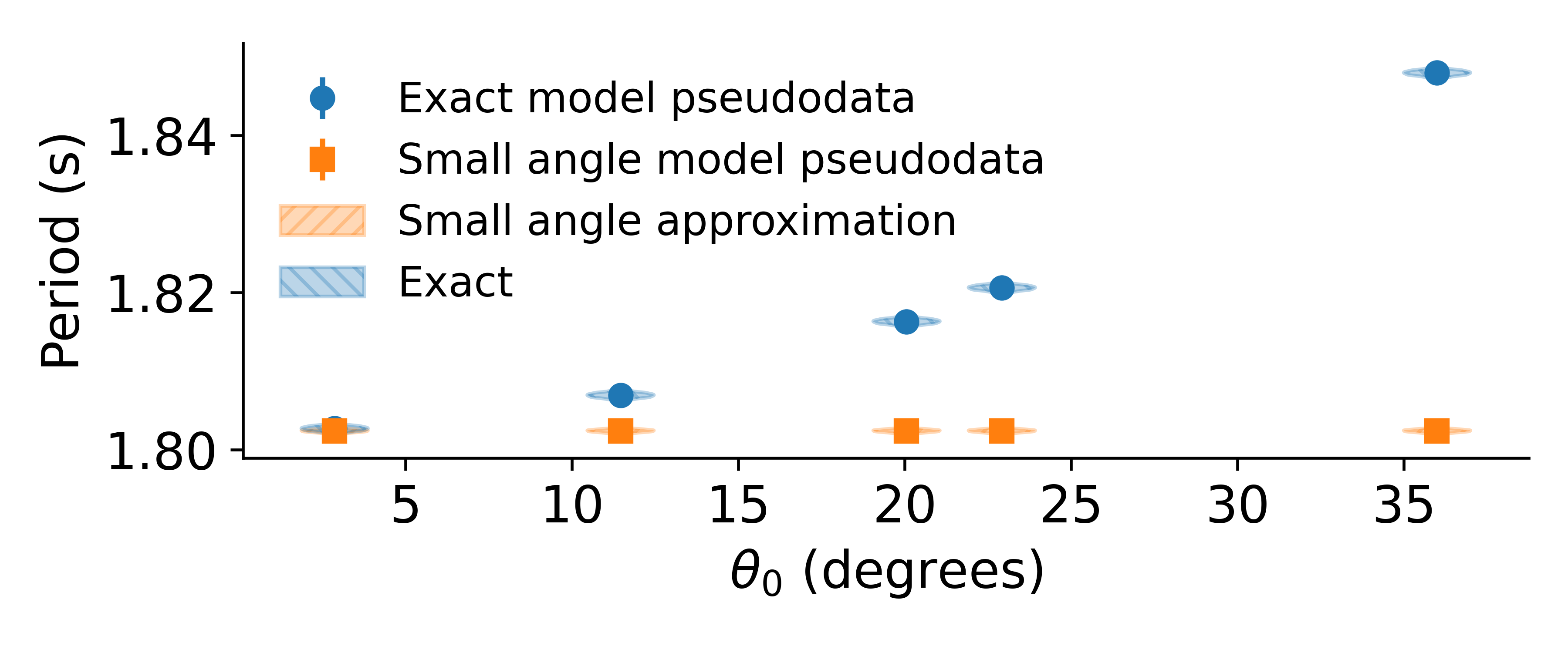}
        \caption{(Color online) Violin plot of the posterior predictive distributions for $g$ from model validation. The posterior distributions are those shown in Fig.~\ref{fig:closureposteriors}. Tails of the posterior correspond to tails in the posterior predictive distribution. }
        \label{fig:closure-predictives}
    \end{figure}

    Just as prior predictive checks were performed, it is important to perform posterior predictive checks to see if the knowledge of $g$ posterior to comparison with data is consistent with the data. 
    Predictions are made by taking draws from the posterior distribution and evaluating the model to form the \emph{posterior predictive distribution} in  Fig.~\ref{fig:closure-predictives}. Both models clearly produce predictions consistent with their respective validation data. 
    A mismatched posterior predictive distribution can indicate an issue in the inference, the MCMC, or 
    may reveal model deficiencies useful for improving the model. For example, the posterior predictive distributions may reproduce some -- but not all -- data, indicating missing physics. 
    Where the models are calibrated to data generated by themselves, both are able to recover the known input (Fig.~\ref{fig:closureposteriors}) and reproduce pseudodata (Fig.~\ref{fig:closure-predictives}). 
    The distribution of model predictions are shown with violin plots, which show distributional information. In violin plots, the width of the regions corresponds to the probability density of the data and is able to show more substructure than a box-and-whisker plot. 
    
	\subsection{Step 4: Inference with data}
    
    Only after the prior and model self-consistency have been established is it appropriate to systematically compare the models to real data. This is because model mis-specification and missing features could be identified and rectified before extracting physics conclusions. 
    
    Measurements were made with a pendulum of length $L=0.807 \pm 0.0005$ m and a cylindrical bob of mass $10 \pm 0.05 g$ with times recorded using a photodiode connected to a Pasco 850 universal interface Model UI-5000. The measurements of initial angular displacement $\theta_0$ were made by a digital protractor in degrees with hundredth-of-a-degree precision, corresponding to a $\pm 0.005^\circ$ angular uncertainty. The bob was released from the same location using a stand. 5 measurements of 10 periods were made at each arbitrary angular displacement, consistent with the typical number of measurements performed by students in introductory teaching laboratories \cite{McGill101}. The means of these measurements and associated uncertainties are provided in Table~\ref{tab:data}.
    
    \begin{table}[htb!]
    \centering
    \begin{tabular}{cc}
    \hline $\theta_0$ (deg) $\pm 0.005^\circ$ \hspace{0.1in} & Period (s) $\pm 0.0001$ s \hspace {0.1in}\\
    \hline  6.40 & 1.8028 \\
    11.73 & 1.8056  \\
    16.41 & 1.8113  \\
    19.99 & 1.8152  \\
    23.18 & 1.8196  \\
    32.67 & 1.8339  \\
    \hline
    \end{tabular}
    \caption{Experimental results for the period of the simple pendulum investigated.}
    \label{tab:data}
    \end{table}
    
\begin{figure}[!htbp]
        \centering
        \includegraphics[width=\columnwidth]{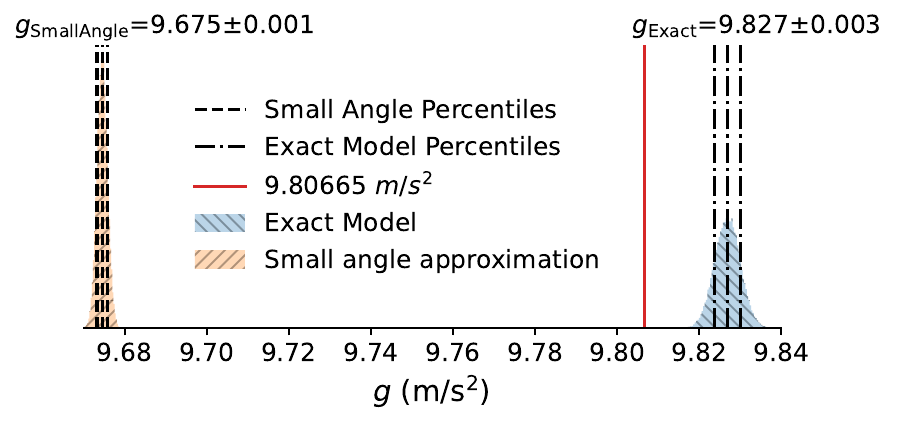}
        \caption{(Color online) Independently normalized posterior distributions for $g$ using the exact formula for the period and the small angle approximation. The standard value \cite{NIST_g} is shown in red and vertical dashed lines denote the 16th, 50th, and 84th percentile of the samples in increasing $g$ with exact values provided above the corresponding distribution. This corresponds to the central value and the 68\% credible interval. Because the two posteriors are independently normalized, it is important to consider each posterior independently and not interpret the relative height of the peaks.}
        \label{fig:posteriors}
    \end{figure}

    The posterior is evaluated numerically using MCMC. The posterior distribution for the gravitational acceleration $g$ is shown in Fig.~\ref{fig:posteriors}. The exact model is able to infer the gravitational acceleration with a relatively precise 68\% credible interval -- $9.818\pm0.003$ $m/s^2$ -- while the small angle approximation infers a result with higher precision -- $9.715 \pm 0.001$ $m/s^2$ -- but sacrifices accuracy with a remarkable $\sim 0.1$ $m/s^2$ bias from the standard value 9.80665 $m/s^2$ \cite{NIST_g} and from 9.80636 $m/s^2$, the best-available model calculation of $g$ for the experimental apparatus' location \cite{EGM2008}. 
    While in Fig.~\ref{fig:closureposteriors}, both models were self consistent and recovered results within expectations, one can clearly see in Fig.~\ref{fig:posteriors} that the small angle approximation's posterior is less consistent with expectations than that of the exact expression for the period.
    
    Increased precision at the cost of bias is commonly known as the ``bias-variance tradeoff".  In the bias-variance tradeoff, a simple model underfits the data and produces unrealistically tight constraint with a bias while a complex model overfits the data and can produce an accurate estimate with excessive variance. Addressing this tradeoff is a focus of model development and many techniques in statistical learning \cite{neal2019biasvariance}. While the reference value is not within the 68\% credible region of the exact model, the bias is greatly reduced and the major features of the data are successfully reproduced. 
    
    Similarly, the bias-variance tradeoff influences the relative precision of the real data posterior versus that of the model validation. Because the models are both less-well suited to the real data, there is both a cost in terms of bias and in misplaced confidence in the estimate. This could be resolved through incorporating a likelihood more robust to outliers and model mismatch, such as the Student's t-likelihood.  
    
    The source of the difference in precision is the same as in the pseudodata comparisons in Fig.~\ref{fig:closureposteriors}. The small angle approximation does not incorporate angular displacement and thus does not incorporate the uncertainty on $\theta_0$ while the exact formula for the period does. 

	\subsection{Step 5: Posterior predictive checks}
	
\begin{figure}[!htbp]
        \centering
        \includegraphics[width=\columnwidth]{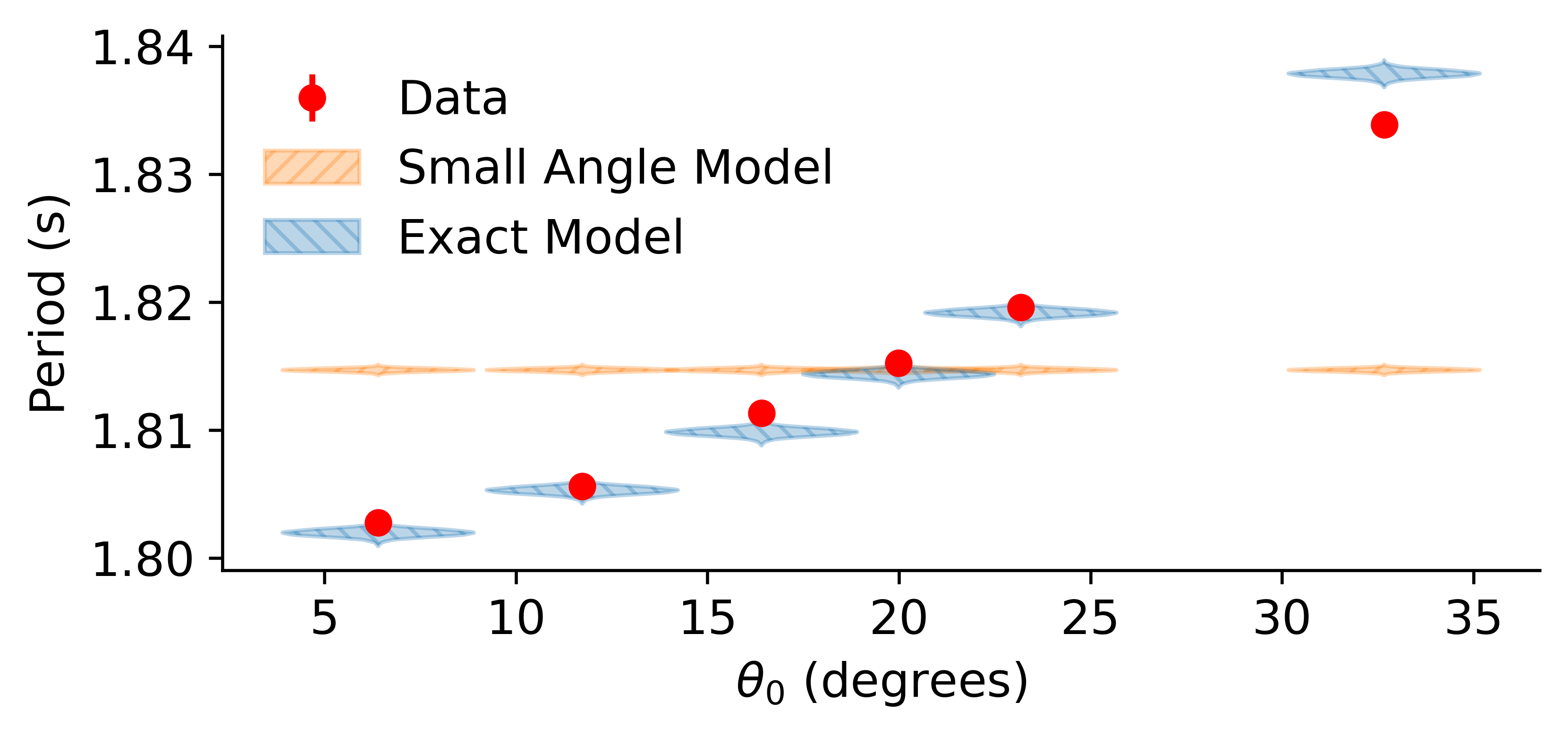}
        \caption{(Color online) Violin plot of the posterior predictive distributions for $g$. The posterior distributions are those shown in Fig.~\ref{fig:posteriors}.}
        \label{fig:predictives}
    \end{figure}

	Posterior predictive distributions can once again elucidate features of the predictions and provide insights into why one model may fail while the other succeeds. The posterior predictive distributions for the models compared to real data are shown in Fig.~\ref{fig:predictives}. The failure of the small angle approximation to reproduce the measured data is striking. As it was shown in Fig.~\ref{fig:closure-predictives} that the model is self-consistent, the only conclusion from Fig.~\ref{fig:predictives} can be that the small angle approximation's lack of $\theta_0$ dependence is the source of its breakdown. As a result, the posterior predictive distributions make the incompleteness of the small angle approximation immediately apparent and interpretable.  
	The predictions from the exact model posterior are constrained around the measurements and are broadly consistent with the measured uncertainty, while the small angle model attempts to match the data by undershooting at large angular displacement and overshooting small displacements.  
	
    This establishes that it is possible to demonstrate bias in calculations of gravitational acceleration introduced by using the small angle approximation while including realistic measurement uncertainties. Through the use of a modeling workflow, it is straightforward to identify and interpret the source of this bias. 

    It is also clear that the exact formula for the period of a simple pendulum contains the majority of relevant physics necessary to describe real data. Features not yet accounted for include the flattening of the data at large displacement in Fig.~\ref{fig:predictives} likely due to drag. Accounting for such effects is beyond the scope of this work.
	
	\section{Model selection} \label{sec:bayesfactor}
    
    It is time to translate the difference between model performance into quantitative guidance using the tools of Bayesian model selection.  
    Bayesian model preference is evaluated using the ratio of two models' Bayes evidences (the denominator of Eq.~\ref{eq:bayes}), called the Bayes factor
    \begin{equation}
        \label{eq:Bayes_factor}
        B_{01} = \frac{p(d | M_0)}{p(d| M_1)}
    \end{equation}
    where models $M_i$ are subscripted 0 and 1. The Bayes factor $B_{01}$ is used to quantify model preference. Hereafter, the exact formula for the period and the small angle approximation are subscripted 0 and 1, respectively. A Bayes factor greater than one represents an increase of support for $M_0$ relative to $M_1$ and directly corresponds to the odds of $M_0$:$M_1$. 
    
    The Bayes factor allows for determination of a preferred model when the result is not immediately obvious. In the simple pendulum, a single observable -- the period -- is considered. However, in more sophisticated experiments, there may be multiple observables considered simultaneously and the comparison becomes more complex (c.f. \cite{SIMS}). While tools such as the $\chi^2$-per-degree-of-freedom exist, these often make implicit assumptions about the shape of the posterior distribution.
    The Bayes factor makes no additional distributional assumptions beyond those explicitly chosen by the priors and the likelihood. 
    A more in-depth discussion of Bayesian model selection may be found in \cite{BayesintheSky}. 
    
    Empirical scales are used to determine when there is weak, moderate, and strong evidence for $M_0$ vs. $M_1$. The Bayes factor is easily interpretable as it gives the direct odds ratio of one model to the other. In this work, the Jeffreys' Scale is used (Table~\ref{tab:jeffreys}). The Jeffreys' Scale is a standard scale for model selection in Bayesian studies \cite{JeffreysScale}. 
    \begin{table}[htb!]
    \centering
    \begin{tabular}{llll}
    \hline$\left|\ln B_{01}\right|$ & Odds & Probability & Strength of evidence \\
    \hline$<1.0$ & $\leq 3: 1$ & $<0.750$ & Inconclusive \\
    1.0 & $\sim 3: 1$ & 0.750 & Weak evidence \\
    2.5 & $\sim 12: 1$ & 0.923 & Moderate evidence \\
    5.0 & $\sim 150: 1$ & 0.993 & Strong evidence \\
    \hline
    \end{tabular}
    \caption{The Jeffreys' Scale, reproduced from \cite{BayesintheSky}.}
    \label{tab:jeffreys}
    \end{table}
    Data are generated for the model selection study as follows. Taking sufficient accurate measurements by hand is unreasonable, involving thousands of measurements of the period, and it has already been demonstrated that the exact expression of the period describes the motion of a pendulum well. Therefore, it is reasonable to generate data using the model, given a set of simple assumptions. It is unreasonable to assume that students would be able to make reliable measurements at increments finer than $1^\circ$ with a meter-scale pendulum or at more than 12 increments in initial angular displacement $\theta_0$.
	Typically, students make fewer measurements in teaching laboratories and this study represents an attempt to find an idealized, maximally-restrictive constraint on initial angular displacement $\theta_0$.
	
	12 proxy data points linearly spaced in $\theta_0$ are generated if the maximum $\theta_0$ is larger than 12$^\circ$. If the maximum $\theta_0$ is less than 12$^\circ$, data are generated in 1$^\circ$ increments. The Bayes factor is then found by calculating the Bayes evidence of each model with respect to the data. It is then possible to determine at what $\rm{Max}[\theta_0]$ the evidence for the exact model becomes weak, moderate, or strong. These data points are generated with the uncertainties commensurate with photodiode, accelerometer, and stopwatch timing. Including these results demonstrates the importance of accounting for uncertainty when offering data-driven guidance and will allow for appropriate implementation of these results in the classroom.
	
	\begin{figure}[!htbp]
        \centering
        \includegraphics[width=\columnwidth]{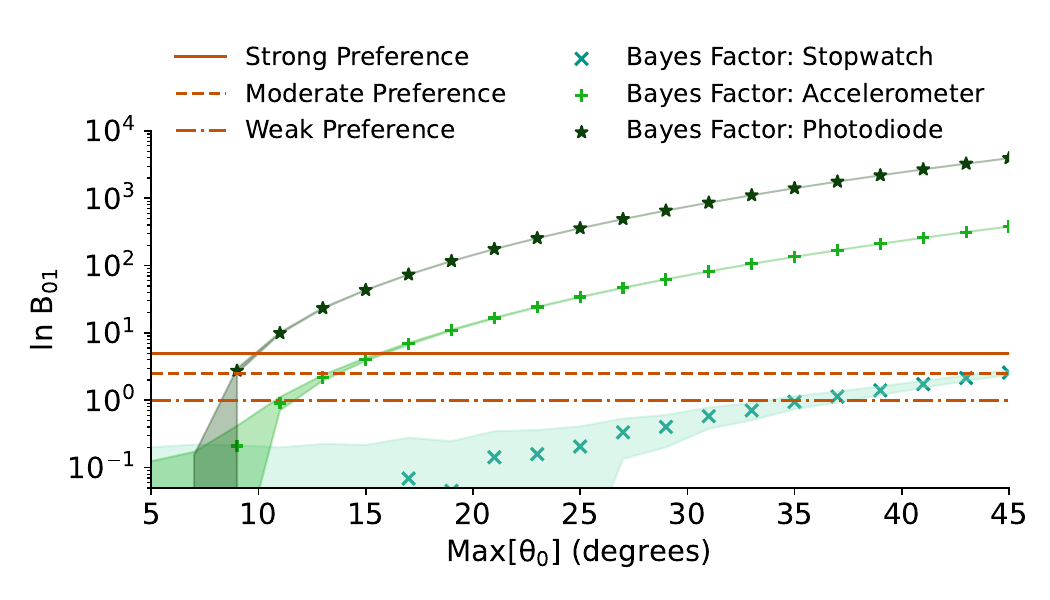}
        \caption{(Color online) $\ln B_{01}$ demonstrating preference for the exact expression for the period of the pendulum over the small angle approximation for a variety of timing methods. Positive values denote preference for the exact model. Horizontal lines denote thresholds for weak, moderate, and strong preference and the filled areas represent the integration uncertainty on the Bayes factors arising from numerical calculation of the Bayes evidence. When timing is imprecise, it becomes difficult to establish model preference.
        }
        \label{fig:bayesfactor}
    \end{figure}
    
    The $\ln$ Bayes factor is shown in Fig.~\ref{fig:bayesfactor}. As anticipated, as the maximum initial angular displacement $\theta_0$ increases, there is progressively stronger preference for the exact expression. The constraint diminishes as the initial angular displacement is restricted to the region where the small angle approximation is increasingly valid. Preference thresholds for the exact calculation at maximum initial angular displacement $\theta_0$ are given in Table~\ref{tab:pref-ex}. The small angle approximation is justified below these model preference thresholds. 
    \begin{table}[htb!]
    \centering
    \begin{tabular}{lcc}
    \hline Strength of evidence & $\left|\ln B_{01}\right|$ & $\theta_0$ (deg)   \\
    \hline Photodiode timing & & \\
    Weak & 1.0 & $8.0^{+0.1}_{-0.2}$  \\
    Moderate & 2.5  & $8.89^{+0.09}_{-0.1}$  \\
    Strong & 5.0  & $9.81 \pm 0.06$ \\
    \hline Accelerometer timing & & \\
    Weak & 1.0  & $11.2 \pm 0.4$  \\
    Moderate & 2.5  & $13.4 ^{+0.2}_{-0.3}$  \\
    Strong & 5.0  & $15.8 \pm 0.2$ \\
    \hline Stopwatch timing & & \\
    Weak & 1.0  & $35 \pm 2$  \\
    Moderate & 2.5 & $44.6 \pm 0.9$  \\
    \hline
    \end{tabular}
    \caption{Model preference thresholds in which the exact formula is preferred over the small angle approximation. Subscript 0 denotes the exact formula and subscript 1 denotes the small angle approximation. Strong preference was not found for stopwatch timing when maximum angular displacement was limited to 45$^\circ$.}
    \label{tab:pref-ex}
    \end{table}
    
    \begin{table}[htb!]
    \centering
    \begin{tabular}{lc}
    \hline Timing Precision & Recommended Restriction\\
    \hline Photodiode ($\pm \sim 0.005$ s) & $10^\circ$  \\
    Accelerometer ($\pm \sim 0.02$ s)& $15^\circ$  \\
    Stopwatch ($\pm \sim \sqrt{2}(0.250)$ s)& 45$^\circ$ \\
    \hline
    \end{tabular}
    \caption{Guidance for restricting initial angular displacement for simple pendula when using the small angular approximation.}
    \end{table}

\section{Conclusions}
    
     This work demonstrates how to approach statistical modeling through a workflow and how to set data-driven constraints. 
     This is the first instance in the literature of modern computational and statistical techniques being applied to the simple pendulum and demonstrates how the modeling procedure followed in research applications may be applied to pedagogical settings. The interpretable failure of the models and how they lead to additional physics insight as a result of this workflow also demonstrate the value that comes with a data analysis workflow. The leveling off of the data immediately suggests the inclusion of drag in the model, allowing for systematic step-by-step model criticism and development. Comparison of the models to data to derive these insights is demonstrative of how data-driven insights can drive physics understanding and inquiry.
     
     The heuristic guidance of constraining simple pendula to $15^\circ$ is shown to be excessively restrictive when timing with the typical precision found in stopwatches, but is reasonable when using a smartphone accelerometer and even overly-permissive when using precise timing methods such as optical timing with photodiodes. The data does not exhibit strong preference for the exact expression for the period with photodiode timing until $9.81 \pm 0.06^\circ$ and with accelerometer timing until $15.8 \pm 0.2^\circ$, while stopwatch timing is insufficiently precise to establish strong preference between the two models at maximum displacements up to $45^\circ$ even under these ideal conditions. This suggests that current guidance of restriction to $\sim 15^\circ$ is appropriate for timing of precision $\pm \mathcal{O}(0.01)$ s, but displacements with more precise timing $\pm \mathcal{O}(0.001)$ s should be restricted to $10^\circ$ while displacements with less precise timing commensurate with stopwatches and human reaction times should not be restricted below $45^\circ$. These restrictions may also be useful in upper-level laboratories with more precise timing or more advanced apparatus such as Kater pendula.
    
    This analysis represents the current state-of-the-art guidance for laboratory manuals and textbooks as it has used realistic best-case uncertainties and advanced statistical techniques to determine when the breakdown of the approximation is actually detectable. Future works should continue with the present guidance, but modify their justification to be data-driven rather than being motivated by a 1\% discrepancy in the small angle approximation. This would both be more directly applicable to instructors and more faithfully communicate the derivation of scientific guidance to students. Additionally, this demonstration sets out a workflow in which students could derive their own guidance in this and other laboratory applications while gaining experience in statistical and scientific modeling.

\section{Acknowledgements}
This work was supported in part by the Natural Sciences and Engineering Research Council of Canada. Thanks are due to C. Gale, N. Provatas, M. Singh, M. Frick, N. Fortier, and R. Yazdi for critical reading of the manuscript and to D. Everett for both a critical reading of the manuscript and detailed discussions on the implementation of Bayesian techniques and their interpretation. R. Turner, P. Movafegh, and S. Biunno provided materials and expertise for obtaining measurements. R. Furnstahl’s resources on Bayesian inference were instrumental in the development of this work. C. Gale, S. Jeon, and K. Ragan provided guidance, useful discussions, and support to this study.

\bibliography{bibliography}

\begin{thebibliography}{26}%
\makeatletter
\providecommand \@ifxundefined [1]{%
 \@ifx{#1\undefined}
}%
\providecommand \@ifnum [1]{%
 \ifnum #1\expandafter \@firstoftwo
 \else \expandafter \@secondoftwo
 \fi
}%
\providecommand \@ifx [1]{%
 \ifx #1\expandafter \@firstoftwo
 \else \expandafter \@secondoftwo
 \fi
}%
\providecommand \natexlab [1]{#1}%
\providecommand \enquote  [1]{``#1''}%
\providecommand \bibnamefont  [1]{#1}%
\providecommand \bibfnamefont [1]{#1}%
\providecommand \citenamefont [1]{#1}%
\providecommand \href@noop [0]{\@secondoftwo}%
\providecommand \href [0]{\begingroup \@sanitize@url \@href}%
\providecommand \@href[1]{\@@startlink{#1}\@@href}%
\providecommand \@@href[1]{\endgroup#1\@@endlink}%
\providecommand \@sanitize@url [0]{\catcode `\\12\catcode `\$12\catcode
  `\&12\catcode `\#12\catcode `\^12\catcode `\_12\catcode `\%12\relax}%
\providecommand \@@startlink[1]{}%
\providecommand \@@endlink[0]{}%
\providecommand \url  [0]{\begingroup\@sanitize@url \@url }%
\providecommand \@url [1]{\endgroup\@href {#1}{\urlprefix }}%
\providecommand \urlprefix  [0]{URL }%
\providecommand \Eprint [0]{\href }%
\providecommand \doibase [0]{http://dx.doi.org/}%
\providecommand \selectlanguage [0]{\@gobble}%
\providecommand \bibinfo  [0]{\@secondoftwo}%
\providecommand \bibfield  [0]{\@secondoftwo}%
\providecommand \translation [1]{[#1]}%
\providecommand \BibitemOpen [0]{}%
\providecommand \bibitemStop [0]{}%
\providecommand \bibitemNoStop [0]{.\EOS\space}%
\providecommand \EOS [0]{\spacefactor3000\relax}%
\providecommand \BibitemShut  [1]{\csname bibitem#1\endcsname}%
\let\auto@bib@innerbib\@empty
%</preamble>
\bibitem [{\citenamefont {Lima}\ and\ \citenamefont
  {Arun}(2006)}]{brazilapprox}%
  \BibitemOpen
  \bibfield  {author} {\bibinfo {author} {\bibfnamefont {F.~M.~S.}\
  \bibnamefont {Lima}}\ and\ \bibinfo {author} {\bibfnamefont {P.}~\bibnamefont
  {Arun}},\ }\bibfield  {title} {\enquote {\bibinfo {title} {An accurate
  formula for the period of a simple pendulum oscillating beyond the small
  angle regime},}\ }\href {\doibase 10.1119/1.2215616} {\bibfield  {journal}
  {\bibinfo  {journal} {American Journal of Physics}\ }\textbf {\bibinfo
  {volume} {74}},\ \bibinfo {pages} {892--895} (\bibinfo {year} {2006})},\
  \bibinfo {note} {\url{https://doi.org/10.1119/1.2215616}}\BibitemShut
  {NoStop}%
\bibitem [{\citenamefont {Blais}(2020)}]{Modelcomparison}%
  \BibitemOpen
  \bibfield  {author} {\bibinfo {author} {\bibfnamefont {Brian~S.}\
  \bibnamefont {Blais}},\ }\bibfield  {title} {\enquote {\bibinfo {title}
  {Model comparison in the introductory physics laboratory},}\ }\href {\doibase
  10.1119/1.5145420} {\bibfield  {journal} {\bibinfo  {journal} {The Physics
  Teacher}\ }\textbf {\bibinfo {volume} {58}},\ \bibinfo {pages} {209--213}
  (\bibinfo {year} {2020})},\ \bibinfo {note}
  {\url{https://doi.org/10.1119/1.5145420}}\BibitemShut {NoStop}%
\bibitem [{\citenamefont {Giancoli}(2016)}]{giancoli2016physics}%
  \BibitemOpen
  \bibfield  {author} {\bibinfo {author} {\bibfnamefont {Douglas~C}\
  \bibnamefont {Giancoli}},\ }\href@noop {} {\emph {\bibinfo {title} {Physics:
  principles with applications}}}\ (\bibinfo  {publisher} {Boston: Pearson},\
  \bibinfo {year} {2016})\BibitemShut {NoStop}%
\bibitem [{\citenamefont {Hinrichsen}(2020)}]{Hinrichsen_2020}%
  \BibitemOpen
  \bibfield  {author} {\bibinfo {author} {\bibfnamefont {Peter~F}\ \bibnamefont
  {Hinrichsen}},\ }\bibfield  {title} {\enquote {\bibinfo {title} {Review of
  approximate equations for the pendulum period},}\ }\href {\doibase
  10.1088/1361-6404/abad10} {\bibfield  {journal} {\bibinfo  {journal}
  {European Journal of Physics}\ }\textbf {\bibinfo {volume} {42}},\ \bibinfo
  {pages} {015005} (\bibinfo {year} {2020})}\BibitemShut {NoStop}%
\bibitem [{\citenamefont {Smith}\ \emph {et~al.}(2020)\citenamefont {Smith},
  \citenamefont {Ashton}, \citenamefont {Vajpeyi},\ and\ \citenamefont
  {Talbot}}]{10.1093/mnras/staa2483}%
  \BibitemOpen
  \bibfield  {author} {\bibinfo {author} {\bibfnamefont {Rory J~E}\
  \bibnamefont {Smith}}, \bibinfo {author} {\bibfnamefont {Gregory}\
  \bibnamefont {Ashton}}, \bibinfo {author} {\bibfnamefont {Avi}\ \bibnamefont
  {Vajpeyi}}, \ and\ \bibinfo {author} {\bibfnamefont {Colm}\ \bibnamefont
  {Talbot}},\ }\bibfield  {title} {\enquote {\bibinfo {title} {{Massively
  parallel Bayesian inference for transient gravitational-wave astronomy}},}\
  }\href {\doibase 10.1093/mnras/staa2483} {\bibfield  {journal} {\bibinfo
  {journal} {Monthly Notices of the Royal Astronomical Society}\ }\textbf
  {\bibinfo {volume} {498}},\ \bibinfo {pages} {4492--4502} (\bibinfo {year}
  {2020})},\ \bibinfo {note}
  {\url{https://academic.oup.com/mnras/article-pdf/498/3/4492/33798799/staa2483.pdf}}\BibitemShut
  {NoStop}%
\bibitem [{\citenamefont {Everett}\ \emph {et~al.}(2021)\citenamefont {Everett}
  \emph {et~al.}}]{SIMS}%
  \BibitemOpen
  \bibfield  {author} {\bibinfo {author} {\bibfnamefont {D.}~\bibnamefont
  {Everett}} \emph {et~al.} (\bibinfo {collaboration} {JETSCAPE}),\ }\bibfield
  {title} {\enquote {\bibinfo {title} {{Multisystem Bayesian constraints on the
  transport coefficients of QCD matter}},}\ }\href {\doibase
  10.1103/PhysRevC.103.054904} {\bibfield  {journal} {\bibinfo  {journal}
  {Phys. Rev. C}\ }\textbf {\bibinfo {volume} {103}},\ \bibinfo {pages}
  {054904} (\bibinfo {year} {2021})},\ \Eprint
  {http://arxiv.org/abs/2011.01430} {arXiv:2011.01430 [hep-ph]} \BibitemShut
  {NoStop}%
\bibitem [{\citenamefont {von Toussaint}(2011)}]{bayesinphysics}%
  \BibitemOpen
  \bibfield  {author} {\bibinfo {author} {\bibfnamefont {Udo}\ \bibnamefont
  {von Toussaint}},\ }\bibfield  {title} {\enquote {\bibinfo {title} {Bayesian
  inference in physics},}\ }\href {\doibase 10.1103/RevModPhys.83.943}
  {\bibfield  {journal} {\bibinfo  {journal} {Rev. Mod. Phys.}\ }\textbf
  {\bibinfo {volume} {83}},\ \bibinfo {pages} {943--999} (\bibinfo {year}
  {2011})}\BibitemShut {NoStop}%
\bibitem [{\citenamefont {Trotta}(2008)}]{BayesintheSky}%
  \BibitemOpen
  \bibfield  {author} {\bibinfo {author} {\bibfnamefont {Roberto}\ \bibnamefont
  {Trotta}},\ }\bibfield  {title} {\enquote {\bibinfo {title} {Bayes in the
  sky: Bayesian inference and model selection in cosmology},}\ }\href {\doibase
  10.1080/00107510802066753} {\bibfield  {journal} {\bibinfo  {journal}
  {Contemporary Physics}\ }\textbf {\bibinfo {volume} {49}},\ \bibinfo {pages}
  {71--104} (\bibinfo {year} {2008})},\ \bibinfo {note}
  {\url{https://doi.org/10.1080/00107510802066753}}\BibitemShut {NoStop}%
\bibitem [{\citenamefont {Sivia}\ and\ \citenamefont
  {Skilling}(2006)}]{sivia2006data}%
  \BibitemOpen
  \bibfield  {author} {\bibinfo {author} {\bibfnamefont {Devinderjit}\
  \bibnamefont {Sivia}}\ and\ \bibinfo {author} {\bibfnamefont {John}\
  \bibnamefont {Skilling}},\ }\href@noop {} {\emph {\bibinfo {title} {Data
  analysis: a Bayesian tutorial}}}\ (\bibinfo  {publisher} {OUP Oxford},\
  \bibinfo {year} {2006})\BibitemShut {NoStop}%
\bibitem [{\citenamefont {Gelman}\ \emph {et~al.}(2013)\citenamefont {Gelman},
  \citenamefont {Carlin}, \citenamefont {Stern}, \citenamefont {Dunson},
  \citenamefont {Vehtari},\ and\ \citenamefont {Rubin}}]{gelman2013bayesian}%
  \BibitemOpen
  \bibfield  {author} {\bibinfo {author} {\bibfnamefont {Andrew}\ \bibnamefont
  {Gelman}}, \bibinfo {author} {\bibfnamefont {John~B}\ \bibnamefont {Carlin}},
  \bibinfo {author} {\bibfnamefont {Hal~S}\ \bibnamefont {Stern}}, \bibinfo
  {author} {\bibfnamefont {David~B}\ \bibnamefont {Dunson}}, \bibinfo {author}
  {\bibfnamefont {Aki}\ \bibnamefont {Vehtari}}, \ and\ \bibinfo {author}
  {\bibfnamefont {Donald~B}\ \bibnamefont {Rubin}},\ }\href@noop {} {\emph
  {\bibinfo {title} {Bayesian data analysis}}}\ (\bibinfo  {publisher} {CRC
  press},\ \bibinfo {year} {2013})\BibitemShut {NoStop}%
\bibitem [{\citenamefont {Gelman}\ \emph {et~al.}(2020)\citenamefont {Gelman},
  \citenamefont {Vehtari}, \citenamefont {Simpson}, \citenamefont {Margossian},
  \citenamefont {Carpenter}, \citenamefont {Yao}, \citenamefont {Kennedy},
  \citenamefont {Gabry}, \citenamefont {Bürkner},\ and\ \citenamefont
  {Modrák}}]{gelman2020bayesian}%
  \BibitemOpen
  \bibfield  {author} {\bibinfo {author} {\bibfnamefont {Andrew}\ \bibnamefont
  {Gelman}}, \bibinfo {author} {\bibfnamefont {Aki}\ \bibnamefont {Vehtari}},
  \bibinfo {author} {\bibfnamefont {Daniel}\ \bibnamefont {Simpson}}, \bibinfo
  {author} {\bibfnamefont {Charles~C.}\ \bibnamefont {Margossian}}, \bibinfo
  {author} {\bibfnamefont {Bob}\ \bibnamefont {Carpenter}}, \bibinfo {author}
  {\bibfnamefont {Yuling}\ \bibnamefont {Yao}}, \bibinfo {author}
  {\bibfnamefont {Lauren}\ \bibnamefont {Kennedy}}, \bibinfo {author}
  {\bibfnamefont {Jonah}\ \bibnamefont {Gabry}}, \bibinfo {author}
  {\bibfnamefont {Paul-Christian}\ \bibnamefont {Bürkner}}, \ and\ \bibinfo
  {author} {\bibfnamefont {Martin}\ \bibnamefont {Modrák}},\ }\href@noop {}
  {\enquote {\bibinfo {title} {Bayesian workflow},}\ } (\bibinfo {year}
  {2020}),\ \Eprint {http://arxiv.org/abs/2011.01808} {arXiv:2011.01808
  [stat.ME]} \BibitemShut {NoStop}%
\bibitem [{\citenamefont {Betancourt}(2019)}]{betancourtfalling}%
  \BibitemOpen
  \bibfield  {author} {\bibinfo {author} {\bibfnamefont {M.}~\bibnamefont
  {Betancourt}},\ }\href
  {https://betanalpha.github.io/assets/case_studies/falling.html} {\enquote
  {\bibinfo {title} {Falling (in love with principled modeling)},}\ } (\bibinfo
  {year} {2019}),\ \bibinfo {note}
  {\url{betanalpha.github.io/assets/case\_studies/falling.html}}\BibitemShut
  {NoStop}%
\bibitem [{\citenamefont {Ku}\ \emph {et~al.}(1966)\citenamefont {Ku} \emph
  {et~al.}}]{ku1966notes}%
  \BibitemOpen
  \bibfield  {author} {\bibinfo {author} {\bibfnamefont {Harry~H}\ \bibnamefont
  {Ku}} \emph {et~al.},\ }\bibfield  {title} {\enquote {\bibinfo {title} {Notes
  on the use of propagation of error formulas},}\ }\href@noop {} {\bibfield
  {journal} {\bibinfo  {journal} {Journal of Research of the National Bureau of
  Standards}\ }\textbf {\bibinfo {volume} {70}},\ \bibinfo {pages} {263--273}
  (\bibinfo {year} {1966})}\BibitemShut {NoStop}%
\bibitem [{\citenamefont {communication~with instructor}(2020)}]{McGill101}%
  \BibitemOpen
  \bibfield  {author} {\bibinfo {author} {\bibfnamefont {Private}\ \bibnamefont
  {communication~with instructor}},\ }\href@noop {} {\enquote {\bibinfo {title}
  {Mc{G}ill {U}niversity {P}hysics 101 lab manual},}\ } (\bibinfo {year}
  {2020})\BibitemShut {NoStop}%
\bibitem [{\citenamefont {Wilkinson}\ and\ \citenamefont
  {Allison}(1989)}]{10.1093/geronj/44.2.P29}%
  \BibitemOpen
  \bibfield  {author} {\bibinfo {author} {\bibfnamefont {Robert~T.}\
  \bibnamefont {Wilkinson}}\ and\ \bibinfo {author} {\bibfnamefont {Sue}\
  \bibnamefont {Allison}},\ }\bibfield  {title} {\enquote {\bibinfo {title}
  {{Age and Simple Reaction Time: Decade Differences for 5,325 Subjects}},}\
  }\href {\doibase 10.1093/geronj/44.2.P29} {\bibfield  {journal} {\bibinfo
  {journal} {Journal of Gerontology}\ }\textbf {\bibinfo {volume} {44}},\
  \bibinfo {pages} {P29--P35} (\bibinfo {year} {1989})}\BibitemShut {NoStop}%
\bibitem [{\citenamefont {team}(2020)}]{stanpriors}%
  \BibitemOpen
  \bibfield  {author} {\bibinfo {author} {\bibfnamefont {Stan~Development}\
  \bibnamefont {team}},\ }\href
  {https://github.com/stan-dev/stan/wiki/Prior-Choice-Recommendations}
  {\enquote {\bibinfo {title} {Prior choice recommendations},}\ } (\bibinfo
  {year} {2020}),\ \bibinfo {note}
  {\url{https://github.com/stan-dev/stan/wiki/Prior-Choice-Recommendations}}\BibitemShut
  {NoStop}%
\bibitem [{\citenamefont {Hirt}\ and\ \citenamefont
  {Featherstone}(2012)}]{moongravity}%
  \BibitemOpen
  \bibfield  {author} {\bibinfo {author} {\bibfnamefont {C.}~\bibnamefont
  {Hirt}}\ and\ \bibinfo {author} {\bibfnamefont {W.E.}\ \bibnamefont
  {Featherstone}},\ }\bibfield  {title} {\enquote {\bibinfo {title} {A
  1.5km-resolution gravity field model of the {M}oon},}\ }\href {\doibase
  https://doi.org/10.1016/j.epsl.2012.02.012} {\bibfield  {journal} {\bibinfo
  {journal} {Earth and Planetary Science Letters}\ }\textbf {\bibinfo {volume}
  {329-330}},\ \bibinfo {pages} {22--30} (\bibinfo {year} {2012})}\BibitemShut
  {NoStop}%
\bibitem [{\citenamefont {{N}{A}{S}{A}}(2021)}]{jupitergravity}%
  \BibitemOpen
  \bibfield  {author} {\bibinfo {author} {\bibnamefont {{N}{A}{S}{A}}},\ }\href
  {https://solarsystem.nasa.gov/planets/jupiter/by-the-numbers/} {\enquote
  {\bibinfo {title} {By the numbers: Jupiter},}\ } (\bibinfo {year} {2021}),\
  \bibinfo {note}
  {\url{https://solarsystem.nasa.gov/planets/jupiter/by-the-numbers/}}\BibitemShut
  {NoStop}%
\bibitem [{\citenamefont {Box}(1980)}]{Box1980SamplingAB}%
  \BibitemOpen
  \bibfield  {author} {\bibinfo {author} {\bibfnamefont {George~EP}\
  \bibnamefont {Box}},\ }\bibfield  {title} {\enquote {\bibinfo {title}
  {Sampling and {B}ayes' inference in scientific modelling and robustness},}\
  }\href@noop {} {\bibfield  {journal} {\bibinfo  {journal} {Journal of the
  Royal Statistical Society: Series A (General)}\ }\textbf {\bibinfo {volume}
  {143}},\ \bibinfo {pages} {383--404} (\bibinfo {year} {1980})}\BibitemShut
  {NoStop}%
\bibitem [{\citenamefont {Thompson}\ and\ \citenamefont
  {Taylor}(2008)}]{NIST_g}%
  \BibitemOpen
  \bibfield  {author} {\bibinfo {author} {\bibfnamefont {Ambler}\ \bibnamefont
  {Thompson}}\ and\ \bibinfo {author} {\bibfnamefont {Barry~N.}\ \bibnamefont
  {Taylor}},\ }\bibfield  {title} {\enquote {\bibinfo {title} {Guide for the
  use of the international system of units ({S}{I})},}\ }\href
  {https://physics.nist.gov/cuu/pdf/sp811.pdf} {\bibfield  {journal} {\bibinfo
  {journal} {NIST {S}pecial {P}ublication}\ }\textbf {\bibinfo {volume} {811}}
  (\bibinfo {year} {2008})},\ \bibinfo {note}
  {\url{https://physics.nist.gov/cuu/pdf/sp811.pdf}}\BibitemShut {NoStop}%
\bibitem [{\citenamefont {Vousden}\ \emph {et~al.}(2015)\citenamefont
  {Vousden}, \citenamefont {Farr},\ and\ \citenamefont {Mandel}}]{ptemcee}%
  \BibitemOpen
  \bibfield  {author} {\bibinfo {author} {\bibfnamefont {W.~D.}\ \bibnamefont
  {Vousden}}, \bibinfo {author} {\bibfnamefont {W.~M.}\ \bibnamefont {Farr}}, \
  and\ \bibinfo {author} {\bibfnamefont {I.}~\bibnamefont {Mandel}},\
  }\bibfield  {title} {\enquote {\bibinfo {title} {{Dynamic temperature
  selection for parallel tempering in Markov chain Monte Carlo simulations}},}\
  }\href {\doibase 10.1093/mnras/stv2422} {\bibfield  {journal} {\bibinfo
  {journal} {Monthly Notices of the Royal Astronomical Society}\ }\textbf
  {\bibinfo {volume} {455}},\ \bibinfo {pages} {1919--1937} (\bibinfo {year}
  {2015})},\ \bibinfo {note}
  {\url{https://academic.oup.com/mnras/article-pdf/455/2/1919/18514064/stv2422.pdf}}\BibitemShut
  {NoStop}%
\bibitem [{pte()}]{ptemcee-code}%
  \BibitemOpen
  \href@noop {} {}\bibinfo {howpublished}
  {\url{https://github.com/willvousden/ptemcee}}\BibitemShut {NoStop}%
\bibitem [{\citenamefont {Foreman-Mackey}\ \emph {et~al.}(2013)\citenamefont
  {Foreman-Mackey}, \citenamefont {Hogg}, \citenamefont {Lang},\ and\
  \citenamefont {Goodman}}]{emcee}%
  \BibitemOpen
  \bibfield  {author} {\bibinfo {author} {\bibfnamefont {Daniel}\ \bibnamefont
  {Foreman-Mackey}}, \bibinfo {author} {\bibfnamefont {David~W.}\ \bibnamefont
  {Hogg}}, \bibinfo {author} {\bibfnamefont {Dustin}\ \bibnamefont {Lang}}, \
  and\ \bibinfo {author} {\bibfnamefont {Jonathan}\ \bibnamefont {Goodman}},\
  }\bibfield  {title} {\enquote {\bibinfo {title} {emcee: The {MCMC} hammer},}\
  }\href {\doibase 10.1086/670067} {\bibfield  {journal} {\bibinfo  {journal}
  {Publications of the Astronomical Society of the Pacific}\ }\textbf {\bibinfo
  {volume} {125}},\ \bibinfo {pages} {306--312} (\bibinfo {year}
  {2013})}\BibitemShut {NoStop}%
\bibitem [{\citenamefont {Pavlis}\ \emph {et~al.}(2012)\citenamefont {Pavlis},
  \citenamefont {Holmes}, \citenamefont {Kenyon},\ and\ \citenamefont
  {Factor}}]{EGM2008}%
  \BibitemOpen
  \bibfield  {author} {\bibinfo {author} {\bibfnamefont {Nikolaos~K.}\
  \bibnamefont {Pavlis}}, \bibinfo {author} {\bibfnamefont {Simon~A.}\
  \bibnamefont {Holmes}}, \bibinfo {author} {\bibfnamefont {Steve~C.}\
  \bibnamefont {Kenyon}}, \ and\ \bibinfo {author} {\bibfnamefont {John~K.}\
  \bibnamefont {Factor}},\ }\bibfield  {title} {\enquote {\bibinfo {title} {The
  development and evaluation of the earth gravitational model 2008
  (egm2008)},}\ }\href {\doibase https://doi.org/10.1029/2011JB008916}
  {\bibfield  {journal} {\bibinfo  {journal} {Journal of Geophysical Research:
  Solid Earth}\ }\textbf {\bibinfo {volume} {117}} (\bibinfo {year} {2012}),\
  https://doi.org/10.1029/2011JB008916}\BibitemShut {NoStop}%
\bibitem [{\citenamefont {Neal}(2019)}]{neal2019biasvariance}%
  \BibitemOpen
  \bibfield  {author} {\bibinfo {author} {\bibfnamefont {Brady}\ \bibnamefont
  {Neal}},\ }\href@noop {} {\enquote {\bibinfo {title} {On the bias-variance
  tradeoff: Textbooks need an update},}\ } (\bibinfo {year} {2019}),\ \Eprint
  {http://arxiv.org/abs/1912.08286} {arXiv:1912.08286 [cs.LG]} \BibitemShut
  {NoStop}%
\bibitem [{\citenamefont {Efron}\ \emph {et~al.}(2001)\citenamefont {Efron},
  \citenamefont {Gous}, \citenamefont {Kass}, \citenamefont {Datta},\ and\
  \citenamefont {Lahiri}}]{JeffreysScale}%
  \BibitemOpen
  \bibfield  {author} {\bibinfo {author} {\bibfnamefont {Bradley}\ \bibnamefont
  {Efron}}, \bibinfo {author} {\bibfnamefont {Alan}\ \bibnamefont {Gous}},
  \bibinfo {author} {\bibfnamefont {R.~E.}\ \bibnamefont {Kass}}, \bibinfo
  {author} {\bibfnamefont {G.~S.}\ \bibnamefont {Datta}}, \ and\ \bibinfo
  {author} {\bibfnamefont {P.}~\bibnamefont {Lahiri}},\ }\bibfield  {title}
  {\enquote {\bibinfo {title} {Scales of evidence for model selection: {F}isher
  versus {J}effreys},}\ }\href {http://www.jstor.org/stable/4356166} {\bibfield
   {journal} {\bibinfo  {journal} {Lecture Notes-Monograph Series}\ }\textbf
  {\bibinfo {volume} {38}},\ \bibinfo {pages} {208--256} (\bibinfo {year}
  {2001})}\BibitemShut {NoStop}%
\end{thebibliography}%

\end{document}